\documentclass[aps,prr,reprint,twocolumn,superscriptaddress,floatfix,nofootinbib,amsmath, amssymb]{revtex4-2}
\usepackage{natbib}
\usepackage[caption=false]{subfig}
\usepackage{nicematrix}
\usepackage{physics}
\usepackage{xr-hyper}
\usepackage{hyperref}
\usepackage{float}   
\usepackage{bbold}
\usepackage[utf8]{inputenc}
\usepackage{graphicx}
\usepackage{tabularray}
\usepackage{tikz}
\usepackage{dcolumn}
\usepackage{bm}
\usepackage{url}
\usepackage{textcase}
\usepackage{color}
\usepackage[normalem]{ulem}

\newcommand{\be}{\begin{eqnarray}}
\newcommand{\ee}{\end{eqnarray}}
\newcommand{\ii}{{\bf i}}
\newcommand{\jj}{{\bf j}}

\newcommand{\utrentoafil}{Physics Department, University of Trento, Via Sommarive 14, I-38123 Trento, Italy}
\newcommand{\tifpaafil}{INFN-TIFPA Trento Institute of Fundamental Physics and Applications, Via Sommarive 14, I-38123 Trento, Italy}
\newcommand{\becafil}{INO-CNR Pitaevskii BEC Center, Via Sommarive 14, I-38123 Trento, Italy}

\begin{document}
\title{Statistical Mechanics of  Heteropolymers from Lattice Gauge Theory }
\author{Veronica Panizza}
\affiliation{\utrentoafil}
\affiliation{\becafil}
\affiliation{\tifpaafil}
\author{Alessandro Roggero}
\affiliation{\utrentoafil}
\affiliation{\tifpaafil}
\author{Philipp Hauke}
\affiliation{\utrentoafil}
\affiliation{\becafil}
\affiliation{\tifpaafil}
\author{Pietro Faccioli}
\affiliation{Physics Department of  University Milan-Bicocca and INFN, Piazza della Scienza 3, I-20126 Milan,   Italy.}
\email{pietro.faccioli@unimib.it}

\begin{abstract}
Lattice models are valuable tools to gain insight into the statistical physics of heteropolymers. We rigorously map the partition function of these models into a vacuum expectation value of a $\mathbb{Z}_2$ lattice gauge theory (LGT), with both fermionic and bosonic degrees of freedom. Because the associated path integral expression is not affected by a sign problem, it is amenable to Monte Carlo (MC) sampling in both the sequence and structure space, unlike conventional polymer field theory. 
At the same time, since  the LGT  encoding relies on qubits, it provides a framework for future efforts to capitalize on the development of quantum computing hardware.  
We discuss two  illustrative applications of our formalism: first, we use it to characterize the thermodynamically stable sequences and structures of small heteropolymers consisting of two types of residues. Next, we assess its efficiency to sample ensembles of compact structures, finding that the MC decorrelation time scales only linearly with the chain length.
\end{abstract}
\maketitle

Lattice polymers have been extensively studied to gain qualitative insight into the statistical physics of a wide class of soft matter systems  \cite{Bryngelson2005}. In particular, compact structures of lattice heteropolymers have emerged as paradigmatic simplified models for protein native states.
While efficient algorithms have been developed for dilute polymer ensembles \cite{binder, LP1, LP2},
compact structures of heteropolymers have proven significantly more challenging to sample \cite{PERM, HDILL}: Indeed, growth-based methods become exponentially inefficient, and the low acceptance rate of pivot moves hinders Monte Carlo (MC) schemes.
Combinatorial algorithms derived from graph theory can be more efficient in generating compact structures, particularly Hamiltonian paths \cite{hamiltonianpath1,hamiltonianpath2}, but do not satisfy ergodicity. Recently developed advanced MC schemes overcome this limitation \cite{kineticpath1,kineticpath2}.  However, despite these advances,  generating large uncorrelated compact heteropolymer lattice structures remains challenging.

In recent years, the development of quantum computing hardware has inspired alternative approaches that rely on a binary encoding of lattice polymer configurations \cite{Micheletti2021, Qfold1, Qfold2, Slongo2023}.
The upgrade from classical to quantum encoding involves replacing each binary variable with a two-level quantum system (qubit).  In the classical and the quantum encoding,  the task of sampling the ergodic surface can be formulated as a Quadratic Unconstrained Binary Optimization (QUBO) problem, enabling the characterization of remarkably complex polymer ensembles, such as self-assembled ring melts \cite{Slongo2023}.
Unfortunately, the computational advantage of the QUBO encoding rapidly degrades when it is applied to ensembles that do not contain ring polymers.  In these cases, many new hard constraints and ancillary variables must be introduced to remove rings up to a given size. The number of these interactions and related ancillary variables diverges in the thermodynamic limit, where rings of any size can occur. 
In addition, the QUBO approach cannot be directly applied to sample heteropolymer ensembles, because the viable conformational states are not confined to a ground space.
In principle, this problem may be overcome by resorting to 
 MC schemes. However, naive local trial moves based on randomly flipping binary variables would not preserve the chain's topology, leading to exceedingly high rejection rates. 

In this work, we overcome all these limitations by introducing a new quantum encoding, in which the partition function is mapped into a vacuum expectation value of a quantum field theory with both fermionic and bosonic degrees of freedom. The dynamics of this theory is shaped by its $\mathbb{Z}_2$ gauge symmetry, which expresses the chain's continuity condition and inspires MC moves that preserve the chain's topology. 
By tuning the coupling constants of our LGT, we can vary the ring density or even set it to zero. As we show below,  this scheme is particularly apt to sample compact polymer configurations.
Furthermore, this LGT can be naturally extended to study heteropolymers by adding  chemistry-specific soft interactions. Finally, this encoding relies on qubits, so it  provides a suitable framework to develop future  quantum computing algorithms (see Sec.~\ref{sec:SI:status_quantum_simulation_LGTs} of the Supplementary Material (SM) \cite{supplementary_material} for an introductory survey on quantum computing application to LGTs).

\begin{figure}[t!]
\centering\includegraphics[width=0.95\linewidth]{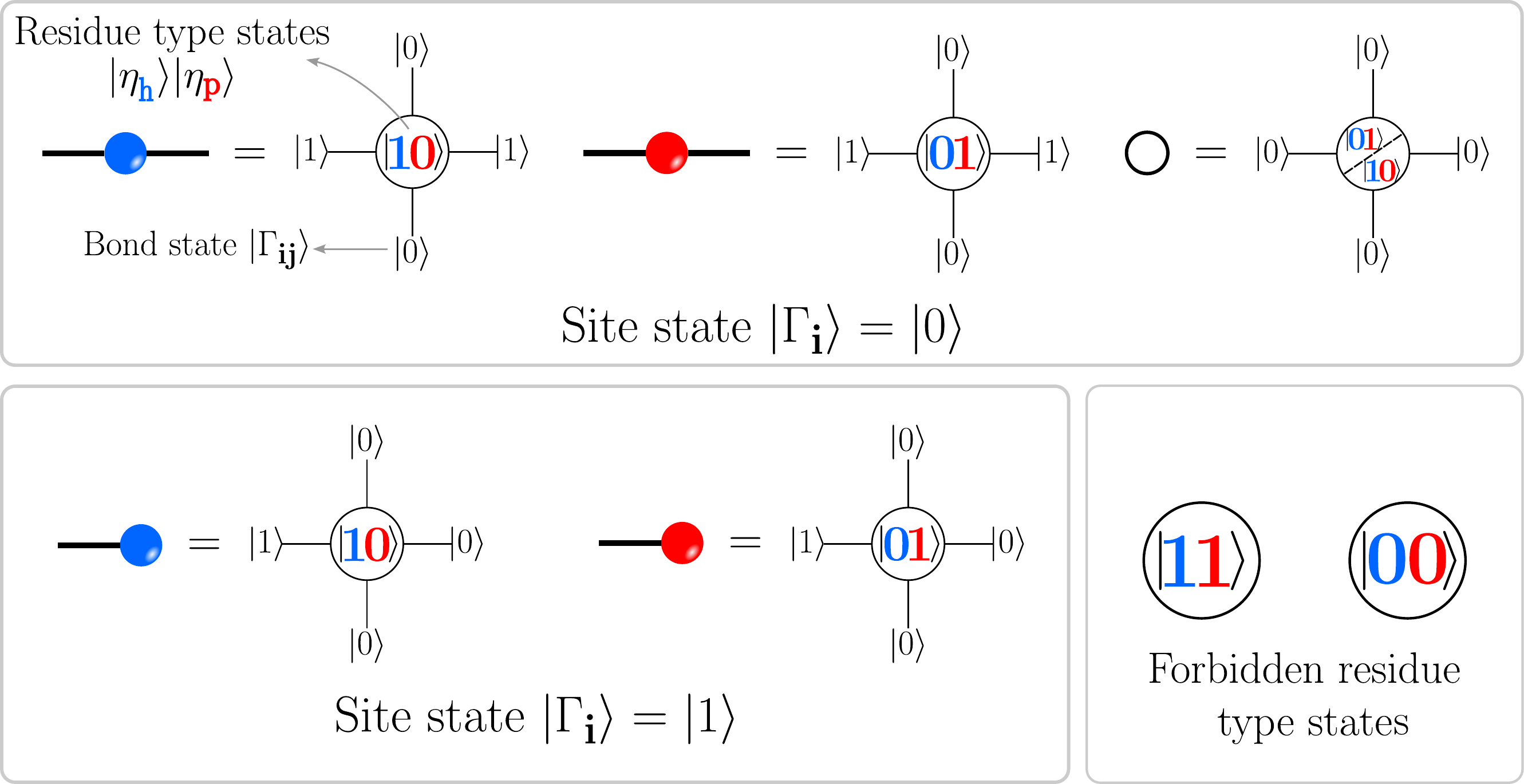}
    \caption{Our quantum encoding, illustrated for the so-called HP model, where residues are grouped into hydrophobic and polar. 
    The colored circles highlight the sites connected to active bonds.  }
    \label{fig:introductory_panel}
\end{figure}

{\bf From polymers to LGT: }  To establish our mapping, we begin by assigning a qubit state $|\Gamma_{\ii\jj}\rangle$ to each edge connecting \emph{neighboring} lattice sites $\ii$ and $\jj$ (see Fig.~\ref{fig:introductory_panel}) and a qubit state $|\Gamma_\ii\rangle$ to each lattice site $\ii$.

The presence (absence) of a covalent bond between residues located at neighboring sites $\ii$ and $\jj$ is signaled by $\ket{\Gamma_{\ii \jj}}=\ket{1} (\ket{0})$, which we call \emph{active} (\emph{inactive}) bond state. Conversely, a chain terminal occupies site $\ii$ if $\ket{\Gamma_\ii}=\ket{1}$. 
We also define the bond operator $\hat{B}_{\ii\jj} \equiv \ketbra{1}$ (projector on active bonds) and the site operator $\hat g_\ii = \ketbra{1}{1}$ (projector on occupied sites). 

We now consider  the tensor product states of all the qubits,  
$
|\Psi\rangle = \left( \prod_{\otimes \langle \ii, \jj\rangle }  \ket{\Gamma_{\ii\jj}}\right) \otimes \prod_{\ii}\ket{\Gamma_\ii}$, constituting our computational basis. 
In a subset of these states,  the qubits in the $\ket{1}$ state align to form continuous paths in the $d$-dimensional lattice. We now show that these so-called \emph{polymer states} 
are  solutions of
\be 
\prod_{k=1}^{d} (-1)^{\hat B_{\ii\, \ii + {\bf e}_k} } \, (-1)^{ \hat B_{\ii\, \ii -{\bf e}_k}} |\Psi\rangle = (-1)^{\hat {g}_\ii} \,  |\Psi\rangle,
   \label{eq:gaussop} 
\ee 
where  $\{{\bf e}_k\}_{k=1, \ldots, d}$ is the set of lattice basis vectors. 
 
To prove Eq.~\eqref{eq:gaussop}, we first note that the computational basis states are eigenstates of the operators at the two sides of the equation. The corresponding eigenvalues obey
\be
\prod_{k=1}^{d}    (-1)^{\Gamma_{\ii \, \ii + {\bf e}_k} + \Gamma_{\ii \,\ii - {\bf e}_k}} = (-1)^{\Gamma_\ii}.
\label{flux}
\ee
Equation~\eqref{flux} expresses a flux conservation condition associated with a binary field $\Gamma_{\ii \jj}$ that is ``emitted" and ``absorbed" at the chain endpoints (see also Fig.~\ref{fig:physical_sector_global} in the SM).  To see this, consider first a generic site $\ii$ where $\Gamma_\ii=0$. There, Eq.~(\ref{flux}) is fulfilled only if the number of active bonds connected to $\ii$, given by the sum
\begin{equation}
\rho_{\ii} = \sum_{k=1}^{d}     \left(\Gamma_{\ii \, \ii + {\bf e}_k} + \Gamma_{\ii \,\ii - {\bf e}_k}\right)\;,
\end{equation}
is zero in empty sites and even in occupied sites.  In particular, for the $3$-dimensional cubic lattice it is equal to 2 along the chain's backbone and 4 or 6 at the intersection point between chains.  Conversely, on the same lattice, the number of active bonds attached to the sites with eigenvalue $\Gamma_\ii=1$ (chain endpoints) is equal to 1, or to 3 and 5 if chain intersections occur at the endpoint. A graphical representation of the analogous constraints for a $2$-d square lattice is reported in Fig.~\ref{fig:physical_sector} of the SM.   

The continuity Eq.~\eqref{eq:gaussop} is in fact the Gauss' law of the $\mathbb{Z}_2$ LGT
 introduced by Wegner and Kogout in Ref.~\cite{Kogut1979introduction}, with  $\hat B_{\ii \jj}$  playing the role of the electric operator and $\hat g_\ii$ that of the topological charge operator.  
Inspired by this connection, we also introduce a so-called \emph{chain deformation operator} $\hat C_\square$, which plays the  role of the magnetic operator of the $\mathbb{Z}_2$ LGT, and is defined on each plaquette on the lattice as 
\be
    \hat C_\square =  \prod_{k}\hat X_{k} \qquad \textrm{with} \quad \hat  X = \ketbra{1}{0} +\ketbra{0}{1}\,,      \label{eq:wilson_loop}
\ee
where  $k$  labels the edges defining the given plaquette $\square$.

Realizing the existence of an underlying gauge symmetry is the key to devise a new type of MC algorithm to sample polymer configurations. Indeed, we note that $\hat C_\square$ commutes with the operator on the left-hand side of Eq.~\eqref{eq:gaussop}.  This implies that if  $|\Psi\rangle$ is a polymer state,  then  $\hat C_\square |\Psi\rangle$  is a different (i.e., locally deformed) polymer state. In practice,  $\hat C_\square$ flips the state of all qubits assigned to the edges of a plaquette (see  Fig.~\ref{fig:wilson_loop_action} of the SM). 
Another class of operators that preserve Eq.~\eqref{eq:gaussop} is the chain terminal displacement operator $\hat D_{\ii \jj}$ that acts on the qubits at the neighboring sites $\ii$ and $\jj$  and on the bond qubit connecting them, 
\be     
    \hat{D}_{\ii\jj} = \hat X_{\ii\jj}\, \hat a^\dag_\ii \, \hat a_\jj + \mathrm{h.c.}\,,\label{eq:Ddef}
\ee  
where $\hat a_{\ii} = \ketbra{0}{1}$. 
In a LGT, this term represents the motion of charged matter with a corresponding instantaneous adjustment of electric field lines. 
Two arbitrary polymer configurations with the same number of open chains can be deformed one into the other by a combination of chain deformations and terminal displacements.

Let us further restrict the set of viable tensor product states to retain only \emph{self-avoiding} polymer states. These can be identified with the ground \emph{space} of the Hamiltonian  $\hat H = \hat H_\textrm{Gauss} +  \hat H_\textrm{SA}$, with 
 \be 
\hat{H}_\textrm{Gauss} &=& 
e_0 \sum_\ii\left(\prod_{k=1}^{d} (-1)^{\hat B_{\ii\, \ii + {\bf e}_k} +\hat B_{\ii\, \ii -{\bf e}_k}} - (-1)^{\hat g _\ii}\right)^2, \\
     \hat H_\textrm{SA} &=& e_1 \, \sum_{\ii} \big(\hat{\rho}_\ii + \hat g_\ii  ) (\hat{\rho}_\ii + (\hat g_\ii- 2)\big)\,, \label{eq:self_crossing}
 \ee  
where  $\hat \rho_\ii \equiv \sum_{k=1}^d \left( \hat B_{\ii \,\ii + {\bf e}_k} + \hat B_{\ii \,\ii - {\bf e}_k}\right)$ counts the number of active links at the site $\ii$. If the couplings $e_0$ and $e_1$ are large and $e_0>e_1$, $\hat{H}$ implements two hard constraints:  $\hat H_\textrm{Gauss}$ enforces the Gauss' law in Eq.~\eqref{eq:gaussop}, while $\hat H_\textrm{SA}$ the self-avoidance condition. Indeed, lattice sites where  $\ket{\Gamma_\ii} =\ket{0}$ yield null eigenvalues of $\hat H_\textrm{SA}$ as long as they are either not attached to any active bond (empty sites) or they are linked to exactly two active bonds (backbone sites). 
Lattice sites where $\ket{\Gamma_\ii}=\ket{1}$ avoid an energetic penalty if they are attached to exactly 1 active bond, thus forming one of the chain's endpoints. 

Without loss of generality, in the following we specialize to the case of a single open chain, i.e., we require that only two $|\Gamma_\ii\rangle$ qubits must be in the $\ket{1}$ states (note, at this stage this restriction still allows for an arbitrary number of additional ring polymers). We denote with $\left\{|\Psi_s\rangle\right\}_{s=1, 2, \ldots}$ the numerable set of all self-avoiding polymer states (i.e., the subset of degenerate ground states of $\hat{H}$) and introduce the linear sum $|\Omega\rangle \equiv \sum_s |\Psi_s\rangle$. 
The matrix element $\mathcal{Z}\equiv  \langle \Omega|\Omega\rangle$ counts the total number of distinct self-avoiding configurations and, thus, is identified with the classical (grand-canonical) partition function of a system consisting of a single chain and arbitrary rings. 
Upon inserting the resolutions of the identity in their respective spaces, $\hat{\mathbb{1}}= \sum_{\Gamma_{\ii \jj}} |\Gamma_{\ii \jj}\rangle \langle \Gamma_{\ii \jj} |$  and
$\hat{\mathbb{1}}'= \sum_{\Gamma_\ii } |\Gamma_{\ii }\rangle \langle \Gamma_{\ii}|$, we obtain an explicit representation: 
\be
\mathcal{Z} = 
\sum_{\Gamma} 
\left(\prod_\ii \delta\left(\mathcal{G}_\ii[ \Gamma]\right)\right)\, e^{-\beta H_\mathrm{SA}[\Gamma]}, \quad \beta=(k_BT)^{-1}.\label{PIZ1}
\ee
In this expression,  we have collectively denoted with $\Gamma$  the fields defined by the eigenstates of the bond and chain terminal operators.
In Eq.~(\ref{PIZ1}), we have imposed the Gauss' law via a set of delta-functions for the functions $\mathcal{G}_\ii[\Gamma] \equiv \prod_{k=1}^{d}    (-1)^{\Gamma_{\ii \, \ii + {\bf e}_k} + \Gamma_{\ii \, \ii - {\bf e}_k}} - (-1)^{\Gamma_\ii}$,  while the self-avoiding condition follows from choosing a large coupling $e_1$ in the Hamiltonian $H_\mathrm{SA}$, ensuring $\beta e_1\gg 1$ for all feasible temperatures. The motivation for using two different notations to express these hard constraints is that we can define ergodic trial moves that preserve Gauss' law while we resort to a Metropolis criterion to enforce self-avoidance. 

In the ensemble defined by Eq.~(\ref{PIZ1}), the density of ring polymers and the average length of the open chain is not fixed.  We now show that this limitation  is overcome by coupling the qubits $|\Gamma_{\ii \jj}\rangle$ and $\ket{\Gamma_\ii}$ to ``spinless fermions" with $M$ degenerate flavors, via the tight-binding Hamiltonian
\be 
    \hat H_\mathrm{F} &=& \sum_{\langle \ii\,\jj    \rangle}  \sum_{m=1}^M \hat\psi^{(m)\dag}_{\ii}\hat\psi^{(m)}_{\jj}\,\hat T_{\ii \jj}\,,
    \label{eq:fermionic_hamiltonian_compact_1}\\
   \Hat T_{\ii\jj} &=& \left[\delta_{\ii \jj} \big(m_\mathrm{f} - \bar{g}^2 \hat \rho_\ii\big) - \lambda^2  \hat B_{\ii\jj}\right],
\label{eq:fermionic_hamiltonian_compact_2}
\ee  
where $\hat \psi^{(m)}_{\ii}$ and $\hat \psi^{(m)\dag}_{\ii}$  are   anti-commuting fields. 
The motivation for introducing these fermions into the theory is that the density of rings in the ensemble can be tuned by varying the parameters $m_\mathrm{f}, \bar{g}$, and $\lambda$.

Again,  we consider the matrix element 
$\mathcal{Z}'= \langle \Omega'|\Omega'\rangle$, where
now $|\Omega'\rangle$  denotes the linear sum of all the degenerate ground states of  $\hat{H}'=\hat{H} + \hat{H}_\textrm{F}$. 
The explicit expression  for  $\mathcal{Z}'$ involves a Gaussian path integral over  $M$ Grassmann fields, which we carry out analytically to obtain 
 \be
 \mathcal{Z}' =    \sum_\Gamma  \left(\det T[\Gamma] \right)^M    \prod_\ii \delta(\mathcal{G}_\ii[\Gamma]) \, e^{-\beta H_\textrm{SA}[\Gamma]},
\label{eq:partition_function}
\ee
where $T_{\ii \jj}(\Gamma)=\delta_{\ii\jj} \big(m_\mathrm{f} - \bar{g}^2 \rho_\ii\big) - \lambda^2  \Gamma_{\ii\jj}$.
Appendix~\ref{sec:SM_coupling_constants} of the End Matter, we explicitly show that by setting
$
    m_\mathrm{f} = 4\bar{g}^2 \; \mathrm{and} \; \lambda^2 = \bar{g}^2,
\label{finetune}
$ ring polymers  are completely removed, because the determinant $\det T(\Gamma)$ vanishes identically for all configurations of the  field $\Gamma$  that contain at least one isolated ring.

Once ring polymers have been suppressed, the length of the remaining single chain is fixed by the total number of bonds. Any given average length $\ell_0$ can be set by introducing a \emph{soft} constraint via the additional Hamiltonian term 
$
\hat H_\textrm{L} = c \sum_{\langle \ii \jj\rangle } (\hat B_{\ii \jj}-\overline{\Gamma})^2
$, 
corresponding to the new matrix element $\mathcal{Z}''=  \langle \Omega'| e^{-\beta \hat H_\textrm{L}}|\Omega'\rangle$. 
The parameter $\overline{\Gamma}\in (0,1)$  controls the average propensity of bond qubits to be in the $\ket{1}$ state. Using the idempotence of  $\hat B_{\ii \jj}$, one can recast the  operator $e^{-\beta \hat H_\textrm{L}}$ into the form of a chemical potential term, yielding  $\mathcal{Z}''\propto  \langle \Omega'| e^{\beta \mu \hat L}|\Omega'\rangle$, 
where  
$\hat{L} = \frac{1}{2}\sum_\ii \hat{\rho}_\ii$ 
counts the total number of active bonds and  $\mu = 2c( 2\bar{\Gamma}-1)$. 

The number of degenerate fermion flavors $M$ and the parameter $\mu$ determine the average chain length $\ell_0$ and the size of the length fluctuations. In Appendix~\ref{sec:app:tuning_average_chain_length} of the End Matter, we derive the value of $\mu$ that yields a given desired $\ell_0$, valid in the limit $M\gg 1$: 
  \be
   \mu \, {\simeq}\, \frac{M}{\beta} \bigg(2\ln 2 - \frac{1}{\ell_0+1}\bigg).\label{scaling}
  \ee
  In the same limit, fluctuations around $\ell_0$ are exponentially suppressed.  
  
{\bf Heteropolymers:} 
Let us now  generalize our LGT theory to include soft interactions that depend on the chemical properties of the residues. 
To account for chemical variability, we introduce additional qubits at each lattice site --one for each type of residue-- and a Hamiltonian that encodes non-covalent interactions:
\be 
  && \hspace{-0.8 cm} \hat H_\mathrm{chem}  
    =  \sum_{\ii\jj}\sum_{a,b=1}^{D}  C_{a b}\, \hat \chi_\ii\, \hat \chi_\jj\, (1-\hat B_{\ii\jj}) 
        \hat n^{a}_\ii \, \hat n^{b}_\jj \, \overline{\Delta}_{\ii\jj},
       \label{Echem1}
\ee 
where $D$ is the size of the chemical alphabet, 
$\hat n^a_\ii = \ketbra{1_a}$ is the lattice site occupation operator acting on the qubit of type $a$ located at site $\ii$, and $\hat \chi_\ii \equiv \frac 12 (\hat \rho_\ii + \hat g_\ii)$ is an operator that identifies the sites belonging to the chain.  
With the self-avoiding constraint, we have $\rho_\ii=0$ (site $\ii$ not occupied by the polymer), $\rho_\ii = 1$ (site occupied by the endpoint), and $\rho_\ii = 2$ (site along the backbone), so $\hat \chi_\ii = \hat{\mathbb{1}}$ at the endpoints and along the backbone. 
The factor $\hat \chi_\ii \hat \chi_\jj $  in Eq.~\eqref{Echem1} ensures that the non-covalent interaction Hamiltonian only couples qubits located at lattice sites that are occupied by the chain. 
Furthermore, the term $(1-\hat B_{\ii\jj})$ excludes non-covalent interactions between covalently bound residues. $\overline{\Delta}_{\ii \jj}$ is a decaying function of the distance $|\ii-\jj|$. In a  nearest-neighbor model,  $\overline{\Delta}_{\ii \jj} =1$ if $|\ii- \jj|=1$ and 0 otherwise. $C_{a b}$ is a $D\times D$ matrix that defines the interaction between the residue types. Finally, to set  the relative abundance of the different residue types along the chain, we introduce the term
\be
    \hat H_\mathrm{rel} &=& g' \sum_{a=1}^D\, \sum_\ii\hat\chi_\ii\bigg( \hat n^{a}_\ii  - \nu_a \bigg)^2\,.
    \label{H_rel}
\ee 
This term can also be written as $\hat H_\textrm{rel} = -\sum_a  \mu_a \hat N^a + g'\sum_a\nu_a^2\sum_\ii \hat{\chi}_\ii$, where $\hat N^a= \sum_\ii \hat \chi_\ii\, \hat n^a_\ii$ counts the number of residues of type $a$ in the polymer backbone and  $\mu_a\equiv g'(2\nu_a-1)$ are the associated chemical potentials. 
Since for a single chain $\sum_i \hat{\chi}_\ii=\hat{L}+1$, the second term contributing to $\hat H_\mathrm{rel}$ can be absorbed into the chemical potential term $\sim \mu$ acting on the bond qubits.
After combining all terms, we obtain the final expression for the partition function of a single heteropolymer: 
\be
    &&\hspace{-0.6 cm}
\mathcal{Z}_\textrm{het}= \sum_{\Gamma, \eta}   \prod_\ii \delta[\mathcal{G}_{\ii}(\Gamma)] \delta[\Phi_\ii(\eta)] \, e^{\beta( \mathcal{L}[\eta, \Gamma]-\mathcal{H}[\eta, \Gamma])}.\qquad
\label{eq:G2}
\ee 
The summation $\sum_\eta$ spans the configurations of the $D$ fields  $\eta^a_\ii$, constructed from the eigenvalues of $\hat n^a_\ii$, while
\be
\mathcal{L}[\Gamma, \eta] &\equiv& \mu L + \sum_a \mu_a N_a + M \textrm{Tr}\log T\\
\mathcal{H}[\Gamma, \eta] &\equiv& H_\textrm{SA}  + H_\textrm{chem}.
\ee
$H_\textrm{chem}$ is obtained by replacing  $\hat \Gamma_{\ii \jj}$ and   $\hat n^a_\ii$  with the binary fields $\Gamma_{\ii \jj}$ and $\eta^a_\ii$  in Eq.~\eqref{Echem1}. Setting $\Phi_\ii = (\sum_{a=1}^D \eta^a_\ii -1)^2$, so $\prod_\ii\delta[\Phi_\ii]$ ensures that at any lattice site $\ii$ there is exactly one active residue-type state (see bottom-right panel in Fig.~\ref{fig:introductory_panel}).

Importantly, the theory defined by  Eq.~\eqref{eq:G2} does not allow us, in general, to choose a specific primary sequence.
Nevertheless, it can be used to characterize the thermodynamically stable sequences along with their structures. 
Specific patterns in the sequence may be imposed by additional interactions. For example, co-polymers can be sampled by adding an ``anti-ferromagnetic" interaction, 
$
\hat H_{\textrm{co-pol}}= \sum_{a,b}\sum_{\langle\ii \jj \rangle} \hat n_\ii^a \hat n_\ii^b \hat B_{\ii \jj}
$, while block co-polymers are obtained from the analog ``ferromagnetic" term. 

{\bf MC Algorithm:}  Equation~\eqref{eq:G2} enables us to simultaneously sample the heteropolymer's conformational and chemical space, using a Metropolis MC scheme.  To this end, we first initialize the binary field $\Gamma$ to fulfill all conformational constraints, i.e., to represent an arbitrary self-avoiding polymer configuration. Likewise, the  initial binary  fields $\eta_a$ should obey the hard constraint set by the term $\prod_\ii\delta[\Phi_\ii]$ in Eq.~(\ref{eq:G2}). 
To propose conformational trial moves, we act at a random location with the deformation operator, Eq.~\eqref{eq:wilson_loop}, or with the terminal displacement operator, Eq.~(\ref{eq:Ddef}).
New sequences are proposed by random point-wise mutations of the chemical residue at arbitrary positions in the lattice.  
These trial moves are accepted or rejected according to a Metropolis criterion defined by the probability density $\propto e^{\beta(\mathcal{L}- \mathcal{H})}$.

{\bf Illustrative Application:}
To illustrate our LGT approach to polymer sampling, we choose a minimalistic setting based on the so-called HP  lattice model~\cite{Yue1993,Lau2005}. In this model, the residues are grouped into hydrophobic ($h$) and polar ($p$), with interaction matrix elements $C_{hh} = -1$, $C_{pp} = C_{hp} = 0$. 
For illustration purposes, we simulate a small chain, $\ell_0=9$,  in a 10$\times$10 square lattice with nearest-neighbor interactions. Such two-dimensional lattice models are commonly adopted in biophysical contexts, as they yield surface-to-volume ratios close to that of realistic small globular proteins. 
Moreover, we choose $M=120$, allowing us to use Eq.~\eqref{scaling}  to fix the chemical potential $\mu$ as a function of $\ell_0$ (see Fig.~\ref{fig:chain_length_analysis} in the SM), and we set $\nu_n=\nu_p=0.5$.  Further details about the model, the MC algorithm, and additional numerical parameters are provided in Sec.~\ref{sec:monte_carlo_simulation} of the  SM. 

\begin{figure}[t!]
\centering
\includegraphics[width=0.95\linewidth]{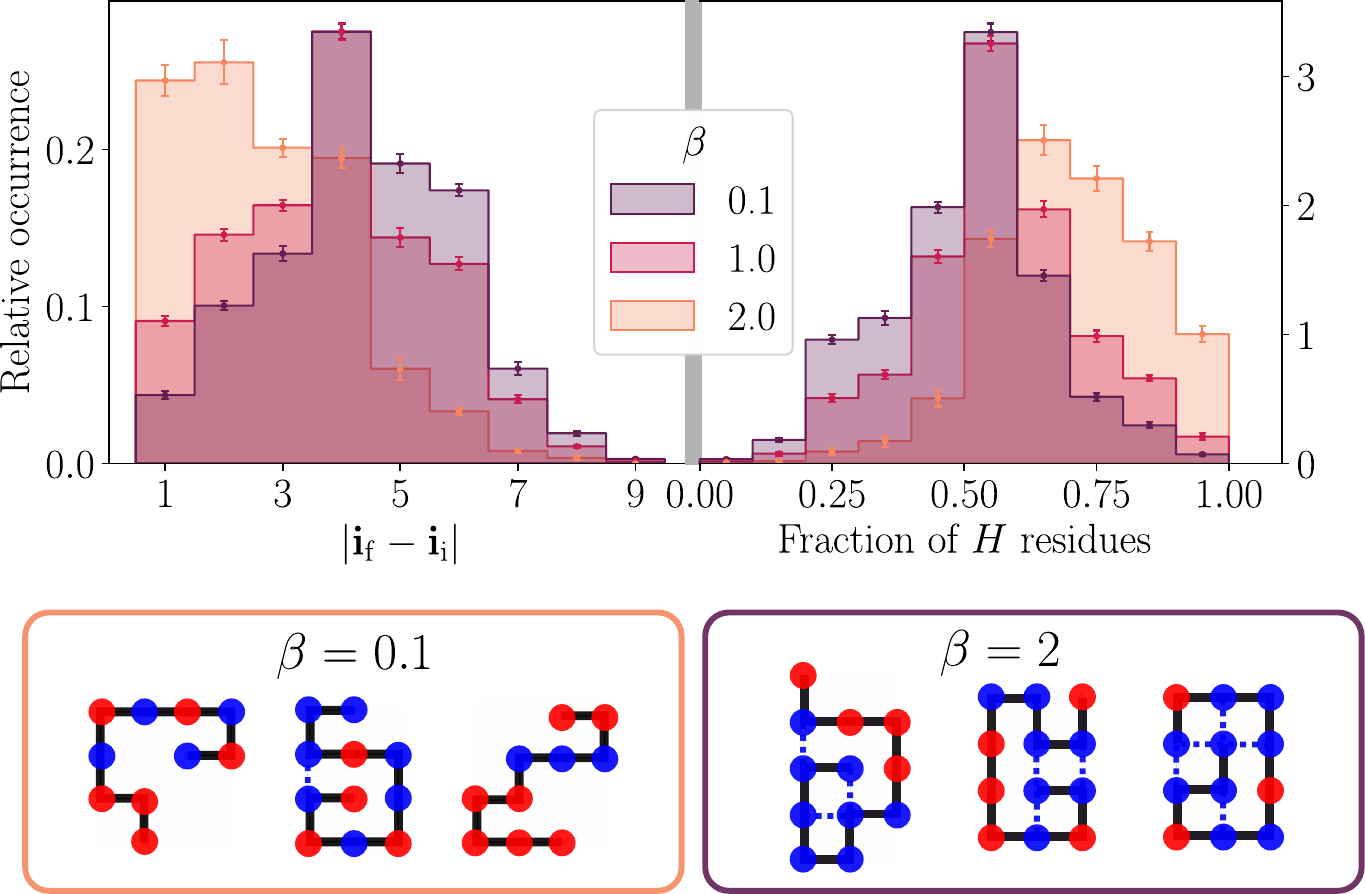}
    \caption{Predictions of the HP model obtained by MC simulation of the LGT, at different inverse temperatures. Upper left: distribution of end-point distance in units of lattice spacing. Upper right: relative occurrence of $h$-type residues for chains with neighboring endpoints.  
    Bottom: samples of representative sequences and structures for two different $\beta$. Dotted lines highlight the network of hydrophobic interactions. }
\label{fig:results_panel}
\end{figure}

In the upper-left panel of Fig.~\ref{fig:results_panel}, we report the end-point distance distributions (in lattice spacing units), which shows that the chain undergoes the hydrophobic collapse in the low-temperature regime. 

Next, we examine how soft interactions promote specific primary sequences. In the upper-right panel of Fig.~\ref{fig:results_panel}, we report the measured relative occurrence of $h$ residues at different temperatures. The latter were arbitrarily set to explore different regimes.    
As can be expected, even when $\mu_h = \mu_p$, the $h$-type residues are more frequent at low temperatures.
Finally, to highlight sequence-structure correlations,  in the bottom panel we report a sample of typical chains generated by our MC. In the low-temperature regime, compact structures are stabilized by the formation of a hydrophobic core, an effect well-known in proteins. 
To quantify this effect, we compare the average number of $h-h$ and $p-p$ non-covalent interactions at $\beta=2$, finding respectively $N_\mathrm{hh} =3.20 \pm 0.03$ and $N_\mathrm{pp} =0.15 \pm 0.01$.
\begin{figure}[t!]
\centering
\includegraphics[width=.8\linewidth]{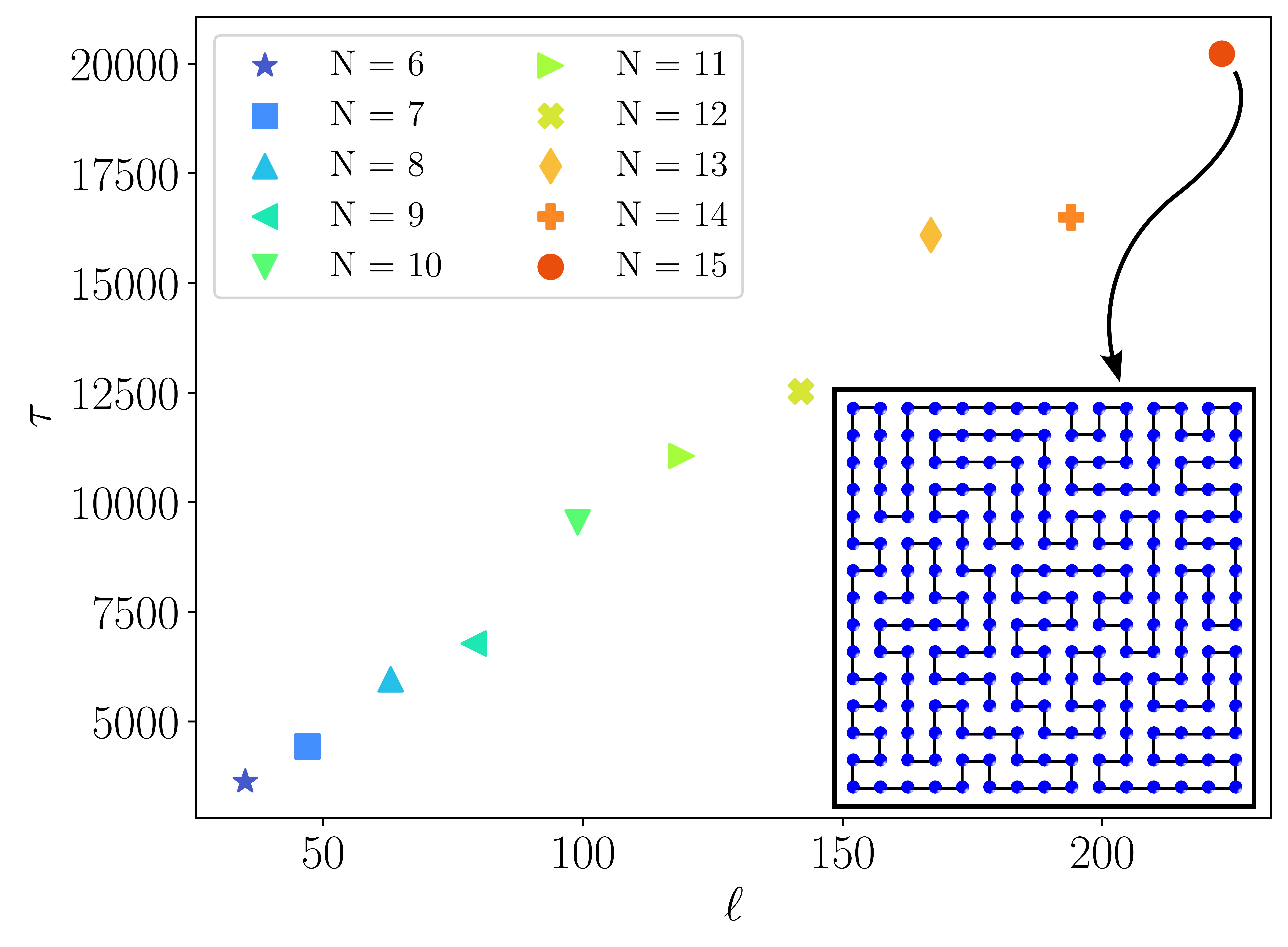}
\caption{Number of MC steps needed to generate decorrelated compact structures ($>98\%$ site occupancy) as a function of the median length $\ell$ (decorrelation time $\tau$), for different two-dimensional lattice sides.  $\tau$ was estimated as twice the exponent that leads the decay of the  autocorrelation function defined in Eq.~\eqref{eq:autocorrelation_formula} of the SM. On the bottom-right corner we show a compact configuration obtained for the largest lattice.}
    \label{fig:scaling}
\end{figure}

{\bf Computational efficiency:} 
The example discussed in Sec.~\ref{sec:SI:sampling_maximally_compact_configurations} of the SM illustrates why our LGT approach is particularly suited to sample compact configurations. To showcase its computational efficiency,  in Fig.~\ref{fig:scaling}, we used our scheme to generate nearly maximally compact configurations ($>98\%$ site occupancy) of a homopolymer chain in a poor solvent. These results show that the MC decorrelation time $\tau$ grows only linearly with the median chain length $\ell$. 

{\bf Discussion and outlook:}  Starting from the pioneering work of Edwards \cite{Edwards1} and DeGennes \cite{degennes1979scaling}, self-consistent field theory methods have  been extensively employed in polymer science\cite{SCPFT, PFTscalingreview}, with applications ranging from material science \cite{SFTmaterial} to biophysics  \cite{Orland2002,GAREL2012}.  
A key limitation of the conventional polymer field theory is that a sign problem prevents the application of stochastic sampling methods. While successful heuristic applications of the  complex Langevin equation have been reported \cite{complexlang1, complexlang2},  a general understanding 
of the conditions under which these schemes converge is still lacking.
In contrast, for even values of $M$ our LGT  does not suffer from a sign problem, and can thus be solved by MC. Furthermore, the  $\mathbb{Z}_2$ gauge symmetry informs conformational moves that are particularly apt to sample compact structures. In the tests we performed on 2-dimensional lattices, the computational cost was found to scale linearly with the median chain length.

LGT was originally developed  to solve quantum chromo-dynamics (QCD) in the strong coupling regime. 
 The present work paves the way for exporting to polymer physics field-theory methods originally developed to investigate  QCD's spectrum and  phase diagram. For example, the vacuum structure and the mass gap in the light-hadron spectrum were shown to be shaped by the spontaneous breaking of chiral symmetry (for a recent review, see, e.g., \cite{Aarts2023}), which involves a transformation of the quark flavors. Using the LGT formulation,  it would be interesting to explore if  the stability and the low entropy of protein native states can also be explained by a spontaneously symmetry breaking. 
Indeed,  previous theoretical \cite{redundant1} and experimental \cite{redundant2} studies have shown that the key properties of the thermodynamics of proteins are preserved by interchanging specific types of amino acids or grouping them into a few effective families, thus providing an approximate ``flavor" symmetry. 

\emph{Acknowledgments}: We acknowledge useful discussions with Henri Orland, Julius Mildenberger, and Luca Spagnoli. We are very grateful to Cristian Micheletti for co-developing the binary encoding of polymer thermodynamics, making important comments,  and pointing out relevant literature.

\bibliographystyle{apsrev4-2}
\bibliography{references}
\clearpage
\onecolumngrid
\section*{End Matter}
\twocolumngrid

\appendix

\section{Setting the density of ring polymers to zero}
\label{sec:SM_coupling_constants}
Without further modifications, the ensemble defined through the partition function in Eq.~\eqref{PIZ1} would contain a chain plus a variable number of loops.
In this appendix, we explicitly prove that, for a suitable choice of parameters, the determinant arising from integrating out Grassmann fields in Eq.~\eqref{eq:partition_function} completely removes the undesired polymer rings from the statistical ensemble. \\

\begin{figure}[b!]
	\centering
	\includegraphics[width=0.7\linewidth]{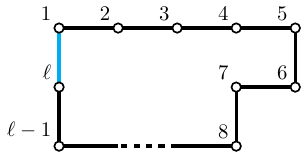}
	\caption{Ring configuration consisting of $\ell$ nodes and $\ell$ edges. 
   In this picture, we omit the representation of the remaining $(N-\ell)$ nodes that do not belong to the loop, whose contribution to the $T(\Gamma)$ matrix in Eq.~\eqref{eq:loopy_configuration_matrix} is embodied by the square sub-matrix $R_{(N-\ell)}$.
   The contribution of the first $\ell$ sites belonging to the loop is reflected by the square sub-matrix $A_{(\ell)}$ specified in Eq.~\eqref{eq:circulant_matrix}, 
   where the entries marked in cyan are associated to the link joining the first node to the $\ell$--th. }
\label{fig:loopy_configuration}
\end{figure}

Let us consider a configuration of the binary field $\Gamma$ associated with an arbitrary polymer configuration containing an  $\ell$--long isolated loop (see   Fig.~\ref{fig:loopy_configuration}). 
To build the Fermion bilinear $T(\Gamma)$ defined in Eq.~\eqref{eq:fermionic_hamiltonian_compact_2},  we adopt the convenient labeling of sites where the first $\ell$ nodes are those belonging to the loop, while the following $(N-\ell)$ identify the remaining vertices left on the lattice, i.e., the ones that are not connected to the loop.
Due to the fact that the loop is \emph{isolated}, the $T(\Gamma)$ matrix in the chosen basis acquires the block diagonal form
\be
    T(\Gamma) = 
    \begin{pNiceArray}{c | c}[margin, columns-width = 1cm]
    A_{(\ell)} & 0 \\ \hline
    0 & R_{(N-\ell)}
    \end{pNiceArray}_{N\times N}\,,
    \label{eq:loopy_configuration_matrix}
\ee
where $A_{(\ell)}$ and $R_{(N-\ell)}$ represent, respectively, the loop and an arbitrary configuration supported on the $(N-\ell)$ remaining vertices.
The sub-matrix $A_{(\ell)}$ is a matrix having the structure
\NiceMatrixOptions{cell-space-limits = 0.15cm}
\be \footnotesize
    A_{(\ell)} = 
    \begin{pNiceArray}{c c c c c}[margin, columns-width=0.2cm]
       m_\mathrm{f} - 2\bar{g}^2& -\lambda^2 & 0 & 0  & \color{cyan}\boldsymbol{-\lambda^2}\color{black}\\
       -\lambda^2 & m_\mathrm{f} - 2\bar{g}^2 & -\lambda^2 & 0 & 0 \\ 
       0& -\lambda^2 & \ddots & \ddots & 0 \\
       0 & \cdot &  \ddots & m_\mathrm{f} - 2\bar{g}^2 & -\lambda^2 \\
       \color{cyan}\boldsymbol{-\lambda^2}\color{black} & 0 & 0 & -\lambda^2 & m_\mathrm{f} - 2\bar{g}^2 
    \end{pNiceArray}_{\ell \times \ell}\hspace{-0.3cm}\,,\nonumber
    \label{eq:circulant_matrix}
\ee
where the entries marked in cyan are related to the connection of the first and the $\ell$--th node, which is highlighted in Fig.~\ref{fig:loopy_configuration} with the same color.
Crucially, these contributions qualify $A_{(\ell)}$ as a circulant matrix having well-studied properties \cite{Gray2006}.

In what follows, we will establish the conditions granting $\det T(\Gamma) = 0$ for all those matrices with the above-mentioned form, 
i.e., those that describe a configuration corresponding to an arbitrarily long isolated loop.
The condition $\det T(\Gamma) = 0$ is equivalent to requiring $T(\Gamma)$ to have at least one \emph{vanishing eigenvalue}. 
Since $T(\Gamma)$ is a block diagonal matrix, a subset of its eigenvalues are those associated with the circulant matrix $A_{(\ell)}$, which can be evaluated exactly (see, e.g.,~\cite{2017Eigenvalues}) to be
\be 
    &\lambda_k = m_\mathrm{f}-2\bar{g}^2 - 2\lambda^2\cos\left(\frac{2\pi k}{\ell}\right)\\
    &\mathrm{for}\, k=1,\dots,\ell\,.\nonumber
\ee
The $\ell$-th eigenvalue of $T(\Gamma)$ vanishes by setting
\be 
m_\mathrm{f} = 2\bar{g}^2 + 2\lambda^2\,.
\label{eq:condition_coupling_constants}
\ee 
Among the many possible choices, this condition can be fulfilled by setting
\be 
m_f = 4\bar{g}^2 \qq{and} \bar{g}^2 = \lambda^2\,,
\label{choice}
\ee 
yielding $\det T(\Gamma)=0$,  for any possible ring number, size, and shape.

\section{\label{sec:app:tuning_average_chain_length}Tuning of average chain length}

\begin{figure}[t!]
	\centering
	\includegraphics[width=0.7\linewidth]{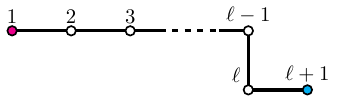}
\caption{Open chain configuration consisting of $\ell$ nodes and $\ell+1$ edges. 
            In this picture, we omit the representation of the remaining $(N-\ell-1)$ nodes that do not belong to the chain, whose contribution to the $T(\Gamma)$ matrix in Eq.~\eqref{eq:open_configuration_matrix} is embodied by the diagonal square sub-matrix $D_{(N-\ell-1)}$.
            The contribution of the first $\ell+1$ sites belonging to the loop is reflected by the square sub-matrix $B_{(\ell+1)}$ specified in Eq.~\eqref{eq:band_matrix}, 
            where the entries marked in magenta and blue are associated to the extremes of the chain (marked with the same colors in the present figure).}
    \label{fig:open_chain_polymer_configuration}
\end{figure}

In this appendix, we show how a suitable choice of parameters permits us to tune the average length of the remaining chain to any desired value. \\

Let us now assume $\bar{g}, m_\textrm{f}$, and $\lambda$ are chosen to obey Eq.~\eqref{choice}, such that we are ensured the absence of any loop. To evaluate Eq.~\eqref{eq:partition_function} of the main text, we require the numerical value of $\det T(\Gamma)$ for the case we are interested in, i.e., binary field configurations $\Gamma$ corresponding to an  open chain of  length $\ell$. 
Remarkably, the structure of the $T(\Gamma)$ matrix does not depend on the specific self-avoiding chain conformation, but only on its length $\ell$ (see Fig.~\ref{fig:open_chain_polymer_configuration}): 
\be
    T(\Gamma) 
    = \begin{pNiceArray}{c | c}[margin]
        B_{(\ell+1)} & 0 \\\hline
        0 & D_{(N-\ell-1)}
    \end{pNiceArray}\,.
    \label{eq:open_configuration_matrix}
\ee
Here, $D_{(N-\ell-1)}$ is a square diagonal matrix with all entries set to $4\bar{g}^2$ and $B_{(\ell+1)}$ takes the form
\be
    B_{(\ell+1)} = 
    \bar{g}^2
    \begin{pNiceArray}{c c c c c}[margin]
    \color{magenta} 3\color{black} & -1 & \cdot & \cdot & \cdot \\ 
          -1 & 2 & -1 & \cdot &\cdot \\
          \cdot      & -1 & \Ddots & \Ddots & \cdot \\
          \cdot & \cdot   & \Ddots & 2 & -1 \\
          \cdot  & \cdot & \cdot & -1 & \color{cyan} 3\color{black}            
    \end{pNiceArray}_{(\ell+1)\times (\ell+1)}\hspace{-0.7cm}\,.
\label{eq:band_matrix}
\ee 
In Sec.~\ref{subsec:SI:determinant_open_chain} of the SM we prove that 
\be 
    \det T(\Gamma) = 4^{N-\ell} \, g^{2N} \, (\ell+1)\,,
    \label{eq:appendix:determinant_open_chain}
\ee 
which enables us to tune the chemical potential $\mu$ acting on link variables to yield any chosen average chain length.  
To this end, we recall that the partition function in Eq.~\eqref{eq:partition_function} 
\be 
    \mathcal{Z}^\prime = \sum_{\Gamma}(\det T(\Gamma))^M \prod_i \delta[\mathcal{G}_i] e^{-\beta H_\mathrm{SA}}\,,
\ee 
in the limit of $\beta e_1 \gg 1$, can we rewritten as
\be 
    \mathcal{Z}^\prime = \sum_{\Gamma_\mathrm{SA}}(\det T(\Gamma_\mathrm{SA}))^M \prod_i \delta[\mathcal{G}_i]\,,
\ee 
where the summation is over the ensemble of self-avoiding lattice states $\Gamma_\mathrm{SA}$. 
By introducing the contribution given by the chemical potential operating on the link variables, we obtain
\be 
    \mathcal{Z}^{\prime\prime} = \sum_{\Gamma_\mathrm{SA}}(\det T(\Gamma_\mathrm{SA}))^M \prod_i\delta[\mathcal{G}_i]\,e^{\beta \mu \ell}\,,
    \label{eq:intermediate_partition_function}
\ee
with $\ell$ being the total number of active links.
If we consider the specific case of a summation over ensembles featuring a single chain $\Gamma_\mathrm{SA}(\ii_\mathrm{i},\ii_\mathrm{j})$ with endpoints $\ii_\mathrm{i}$ and $\ii_\mathrm{f}$ (i.e., lattice sites occupied by topological charges and thus associated to the states $\ket{\Gamma_{\ii_\mathrm{i}}} = \ket{1}$ and $\ket{\Gamma_{\ii_\mathrm{f}}}=\ket{1}$), Eq.~\eqref{eq:intermediate_partition_function} reads
\be 
    \mathcal{Z}^{\prime\prime}_{(1)} = \sum_{\ii_\mathrm{i}, \ii_\mathrm{f}} \sum_{\Gamma_\mathrm{SA}(\ii_\mathrm{i}, \ii_\mathrm{f})}  (\det T(\Gamma_\mathrm{SA}))^M \prod_i\delta[\mathcal{G}_i] e^{\beta\mu\ell}\,.
    \label{eq:one_chain_partition_function}
\ee 

In light of Eq.~\eqref{eq:appendix:determinant_open_chain}, $\det T(\Gamma_\mathrm{SA})$ in Eq.~\eqref{eq:one_chain_partition_function} depends solely on the total length of the chain, making it possible to rewrite Eq.~\eqref{eq:one_chain_partition_function} as
\be
    \mathcal{Z}^{\prime\prime}_{(1)}
    &= \sum_{\ii_\mathrm{i},\ii_\mathrm{f}}\sum_{\ell} n(\ell,\ii_\mathrm{i},\ii_\mathrm{f}) \, \big(\det T(\Gamma)\big)^ M \, e^{\beta\mu \ell} \,,
\label{eq:approx_partition_function_ell}
\ee 
where $n(\ell,\ii_\ii,\ii_\mathrm{f})$ is the number of binary field configurations describing a chain of length $\ell$ having one end-point at the site $\ii_\ii$ and the other at $\ii_\mathrm{f}$.

Using Eq.~\eqref{eq:appendix:determinant_open_chain} we obtain: 
\begin{equation}\begin{split}
    \mathcal{Z}^{\prime\prime}_{(1)}
        &= \sum_{\ii_\mathbf{i},\ii_\mathbf{f}}\sum_{\ell} n(\ell,\ii_\mathbf{i},\ii_\mathrm{f}) \, e^{M \big[2(N-\ell) \ln 2 + N\ln \bar{g}^2 + \ln{(\ell+1)}\big] + \beta \mu \ell}  \\
        & = \sum_{\ell} n(\ell) \, e^{M \big[2(N-\ell) \ln 2 + N\ln \bar{g}^2 + \ln{(\ell+1)}\big] + \beta \mu \ell} \,,
     \label{eq:approx_partition_function_ell_explicit}
\end{split} \end{equation} 
where $n(\ell) = \sum_{\ii_\mathbf{i},\ii_\mathrm{f}}n(\ell,\ii_\mathbf{i},\ii_\mathrm{f})$ is the total number of chains having length $\ell$, irrespectively of the position of the chain's end-points.
To identify the most probable value of the chain length  $\ell_0$ we consider the saddle-point solution of Eq.~\eqref{eq:approx_partition_function_ell_explicit}, i.e.:
\be 
    \left.\partial_\ell \ln n(\ell) \right|_{\ell_0} - M \bigg(2 \ln 2 - \frac{1}{\ell_0 + 1}\bigg) + \beta\mu = 0\,.
    \label{eq:fix_mu_intermediate}
\ee 

Our goal is to solve this equation for $\mu$, for an arbitrarily chosen value of $\ell_0$. 
Unfortunately, computing $\partial_\ell\log n(\ell)$ involves the challenge of estimating the number $n(\ell)$ of independent lattice field states describing polymer configurations of length $\ell$.
Fortunately, this problem can be overcome by considering the regime in which the chemical potential $\mu$ and the number of degenerate fermion fields $M$ are simultaneously large. 
 
With this choice, Eq.~\eqref{eq:fix_mu_intermediate} simplifies to
\be 
    \lim_{M,\mu\to \infty} \frac{\mu}{M}  \simeq   \frac{1}{\beta}\,\bigg(2 \ln 2 - \frac{1}{\ell_0 + 1}\bigg)\,.
     \label{eq:mu_fine_tune}
\ee 
For any (large) choice of $M$,  Eq.~\eqref{eq:mu_fine_tune} allows us to set a value of the chemical potential $\mu$ such that the most frequently sampled chains have length $\ell_0$. 
 
In the large $M$ limit, the distribution of $\ell$ is exponentially peaked around the mean-value $\ell_0$. 
 To see this, we first re-write the partition function of Eq.~\eqref{eq:approx_partition_function_ell_explicit} as
 \be 
    \mathcal{Z} = \sum_\ell e^{-f(\ell)}  \,,
\label{eq:approx_partition_function_saddle_point}
\ee 
with $f(\ell) = - \ln n(\ell) - M \big[ 2(N-\ell)\ln 2 + N\ln \bar{g}^2 + \ln{(\ell+1)}\big]  - \beta\mu\ell$.
Expanding $f(\ell)$ to second order around the saddle-point $\ell_0$, where $\left. \partial_\ell f(\ell)\right|_{\ell_0} = 0$, we  obtain
\be
   f(\ell) 
   & \simeq & f(\ell_0) +  \frac 12 (\ell - \ell_0)^2 \bigg(-\partial^2_\ell \ln n(\ell)\eval_{\ell_0} + \frac{M}{(\ell_0+1)^2}\bigg)\nonumber\\
&&\stackrel{\simeq}{{}_{M, \mu\to \infty}}  f(\ell_0) + \frac{M}{2} \frac{(\ell - \ell_0)^2}{(\ell_0 +1)^2}\,,  
\ee 
Hence, chain length fluctuations around $\ell_0$ are exponentially suppressed in the large $M$ limit.

\onecolumngrid\newpage
\begin{center}
	    \vskip 60pt
	\large \textbf{Supplementary Material} \par
	\vspace{0.4cm}
	\normalsize{Veronica Panizza,$^{1,2,3}$ Alessandro Roggero,$^{1,3}$ Philipp Hauke,$^{1,2,3}$ and Pietro Faccioli$^{4}$}\par
	\vspace{0.15cm}
	\small\textit{$^1$ \utrentoafil}\par
	\small\textit{$^2$ \becafil} \par
	\small\textit{$^3$ \tifpaafil}\par
	\small\textit{$^4$ Physics Department of  University Milan-Bicocca and INFN, Piazza della Scienza 3, I-20126 Milan,   Italy.} 
	\vspace{0.25cm}
\end{center}
	
\twocolumngrid
\renewcommand\thefigure{S\arabic{figure}}  
\renewcommand{\theequation}{S\arabic{equation}}

\setcounter{figure}{0}  
In this Supplementary Material, we provide details about the numerical algorithm and some theoretical results discussed in the main text. In particular:
\begin{enumerate}
    \item We explicitly define the  Monte Carlo (MC) Malgorithm used in the illustrative application, including the numerical values of all the parameters. We also report additional results concerning the autocorrelation time of our simulations and the distribution of chain lengths that we recover by running our algorithm.
    \item We derive Eq.~\eqref{eq:appendix:determinant_open_chain} of the End Matter, which enables us to set an explicit numerical value for the chemical potential $\mu$ in order to sample chains of given average length $l_0$.
    \item We provide one example that illustrates computational advantages of our LGT approach relative to the conventional real-space MC to sample dense lattice polymer ensembles.
    \item We provide additional figures to illustrate the contribution of different terms in the system's Hamiltonian. 
    \item We review recent simulations of lattice gauge theories on quantum devices.
\end{enumerate}

\section{Details about our MC algorithm}
\label{sec:monte_carlo_simulation}

To initialize our MC simulations, we generate the initial configuration of the binary field $\Gamma_{\ii \jj}$ to (i) connect two arbitrarily chosen chain endpoints, (ii) satisfy the chain topology condition---Gauss' law in Eq.~\eqref{flux} of the main text---, and (iii) fulfill the self-avoidance condition. 

To study the NP model employed in the illustrative application,  we randomly initialize the field of binary variables $\eta_\ii^a$ associated with the chemical residue species $a=n$ and $a=p$, in such a way that at each lattice point $\ii$ either $\eta_\ii^n=1$ and $\eta_\ii^p=0$ or $\eta^n_\ii=0$ and   $\eta^p_\ii=1$. 

To generate a new element of the Markov chain, we propose ``backbone deformation" moves, ``end-point diffusion" moves, and ``point-wise monomer mutation" moves, with equal probability.  

\begin{figure}[H]
\centering
\includegraphics{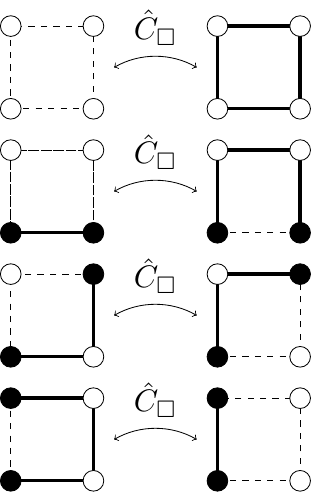}
    \caption{Action of the chain deformation operator in Eq.~\eqref{eq:wilson_loop}. Active and inactive bonds around a plaquette are inverted, while the topological charges remain unaltered.}
    \label{fig:wilson_loop_action}
\end{figure}

To perform backbone deformation moves, we randomly pick a plaquette on the lattice and apply the corresponding chain deformation operator  $\hat C_\square$, defined in  Eq.~\eqref{eq:wilson_loop} of the main text. 
The result of acting with this operator on a plaquette is to switch the value of each bond variable, as shown in Fig.~\ref{fig:wilson_loop_action}. 
To shift the endpoints of the chain, we randomly select one of the two extremes of the chain, $\ii$, and one of the $2d$ oriented directions, $\vb{e}_k$, and we apply the operator $\hat D_{\ii\, \ii+\vb{e}_k}$ defined in Eq.~\eqref{eq:Ddef}, which acts on the backbone configuration as shown in Fig.~\ref{fig:diffusion_operator}.

To perform point-wise monomer mutations, we apply the operator $\hat M_\ii(a)$ that turns the chemical element in site $\ii$ into flavor $a$ at random lattice location site and chemical element. 

The Metropolis acceptance/rejection  criterion for this set of moves is simply
\be 
\hspace{-0.4cm}
    \mathcal{A}\left(\Gamma,\eta \rightarrow \Gamma,\hat M_\ii(a)\,\eta\right) = 
    \min \left(1, \frac{\mathcal{P}(\Gamma,\hat M_\ii(a)\,\eta)}{\mathcal{P}(\Gamma,\eta)} \right)\,, \qquad
\ee 
with $\mathcal{P}(\Gamma,\eta)$ defined as
\be
    \mathcal{P}(\Gamma,\eta)    \propto e^{\mathcal{L}[\Gamma,\eta] - \beta H_\mathrm{chem}[\Gamma,\eta]} \,,
   \label{eq:joint_probability}
\ee 
and 
\begin{equation}
    \begin{split}
    &\mathcal{L}[\Gamma,\eta] = M\Tr\log T[\Gamma] +  \mu \sum_{\ii\jj}\Gamma_{\ii\jj} + \sum_{a,\ii} \mu_a\,  \eta^a_{\ii}  \quad \mathrm{and} \\ 
    &\mathcal{H}[\Gamma,\eta] = H_{SA}[\Gamma] + \sum_{\ii\jj} \sum_{a,b=1}^D C_{ab} \;\chi_\ii\eta^a_\ii\, (1-\Gamma_{\ii\jj}) \, \chi_\jj\eta^b_\jj \; \overline{\Delta}_{\ii\jj}\,. 
    \end{split}
\end{equation}
All the parameters appearing in these equations that we used in our numerics, along with the number and length of the  Markov chains, are reported in Table \ref{tab:numerical_details}. 
We found these to present good performance while reproducing the relevant physics.\\

\renewcommand{\arraystretch}{1.5}
\begin{table}[t!]
    \centering
    \setlength{\tabcolsep}{3mm}
    \begin{tblr}{p{1.5cm}p{4cm}p{2cm}}
    \textbf{Symbol} & \textbf{Definition} & \textbf{Value} \\\hline
    $M$ & Number of fermionic degenerate modes & 120 \\
    $m_\mathrm{f}$ & Mass of fermions & 1\\
    $\beta$ & Inverse temperature & $\{ 0.1,1,2\}$\\
    $e_1$ & Energy scale entering the repulsive term $H_\mathrm{SA}$ (Eq.~\eqref{eq:self_crossing} in the main text) &  $1,000$\\
    -- & Lattice format & $10 \times 10$ \\ 
    $D$ & Number of chemical elements & 2 \\
    $\nu_n = \nu_p$ & Abundance of $n$ and $p$ chemical species & $0.5$ \\
    $\mu_n = \mu_p$& Chemical potential associated with $n$ and $p$ chemical species & $0$ \\
    $\ell_0$ & Expected average chain's length (when no chemical interaction takes place) & 9 \\
    $N_\mathrm{c}$ & Number of polymeric chains generated & 1 \\
    $N_\mathrm{steps}$ & Number of MC steps & $10^6$\\
    $N_\mathrm{mc}$ & Number of MC trajectories for each parameter choice & 10 \\
    $\tilde{\Delta}_{0.1}$ & Decorrelation time when $\beta=0.1$ & $1,800$ \\ 
    $\tilde{\Delta}_{1}$ & Decorrelation time when $\beta=1$ & $1,980$ \\ 
    $\tilde{\Delta}_{2}$ & Decorrelation time when $\beta=2$ & $5,040$ \\ 
    \end{tblr}
    \caption{Parameters adopted in our MC simulations.}
    \label{tab:numerical_details}
\end{table}

\textbf{Autocorrelation  analysis:} The conformational decorrelation time along a   MC trajectory can be estimated from the autocorrelation function $A(\Delta)$, defined as
\begin{equation}
\begin{split}
    A(\Delta) 
    &= \frac{1}{N}\sum_{n=1}^{N} \bigg(\frac{1}{2 N_\mathrm{e}} \sum_{\langle\ii,\,\jj\rangle} \Gamma^{(n)}_{\ii\jj} \Gamma^{(n+\Delta)}_{\ii\jj}\bigg)  + \\ 
    &-\bigg(\frac{1}{N}\sum_{n=1}^{N}\frac{1}{2 N_\mathrm{e}}\sum_{\langle\ii,\,\jj\rangle}\Gamma^{(n)}_{\ii\jj}\bigg)
    \bigg(\frac{1}{N}\sum_{n=1}^N \frac{1}{2 N_\mathrm{e}} \sum_{\langle\ii,\,\jj\rangle}\Gamma^{(n+\Delta)}_{\ii\jj}\bigg)\bigg]\,,
    \label{eq:autocorrelation_formula}
\end{split}\end{equation}
where $N_\mathrm{e}$ is the number of edges in the lattice, $\Gamma^{(n)}_{\ii\jj}$ and $\Gamma^{(n+\Delta)}_{\ii\jj}$ are binary variables sampled at the $n$--th and $(n+\Delta)$--th element of the Markov chain, and the sum $\sum_n$ runs over the MC steps. 
The analogous auto-correlation function used to estimate the decorrelation time of the monomer field is obtained replacing $\Gamma_{\ii \jj}$ with $\Gamma_\ii$

\begin{figure}[t!]
    \centering
    \subfloat{\includegraphics[width=0.9\linewidth]{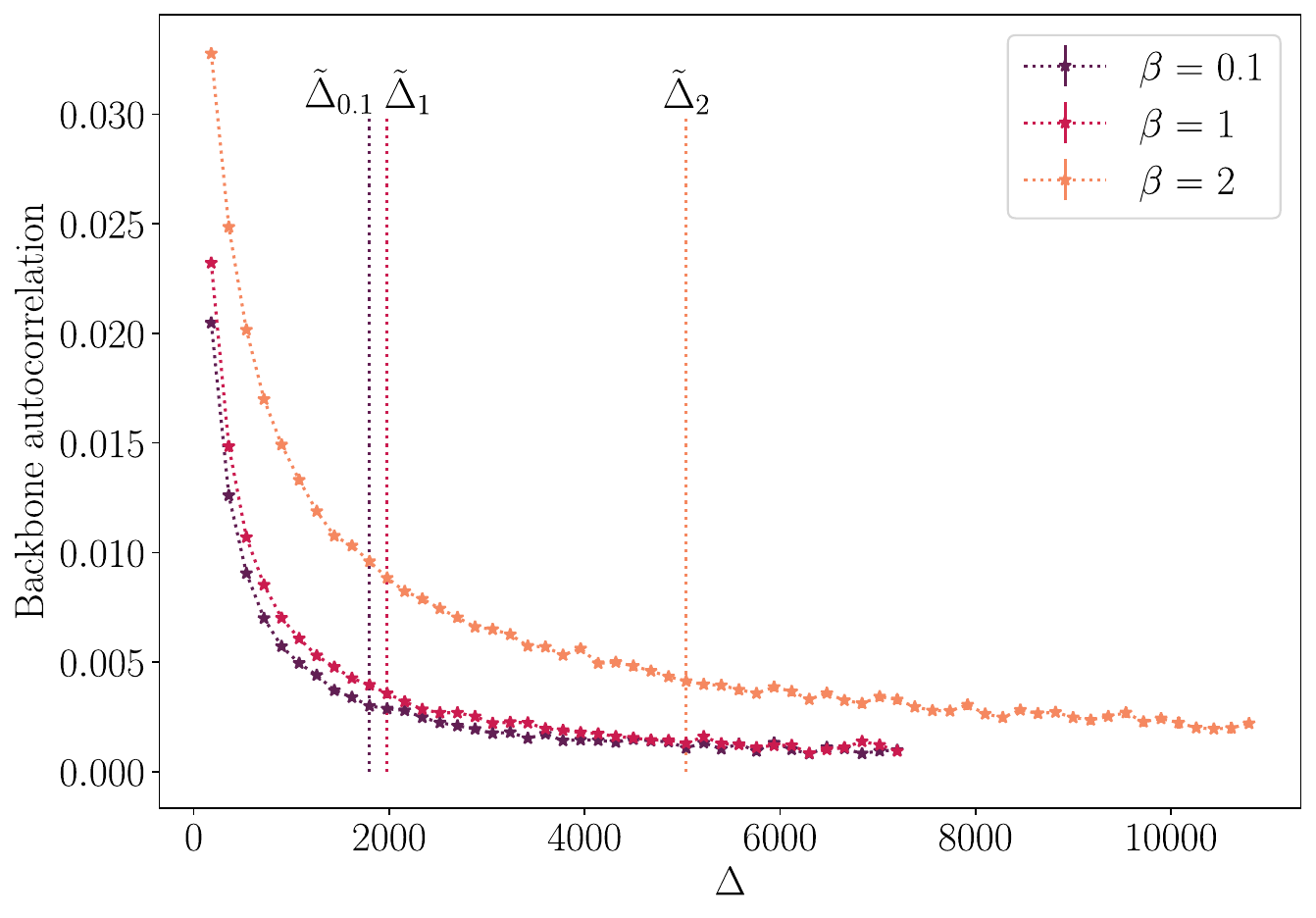}}\\
    \subfloat{\includegraphics[width=0.9\linewidth]{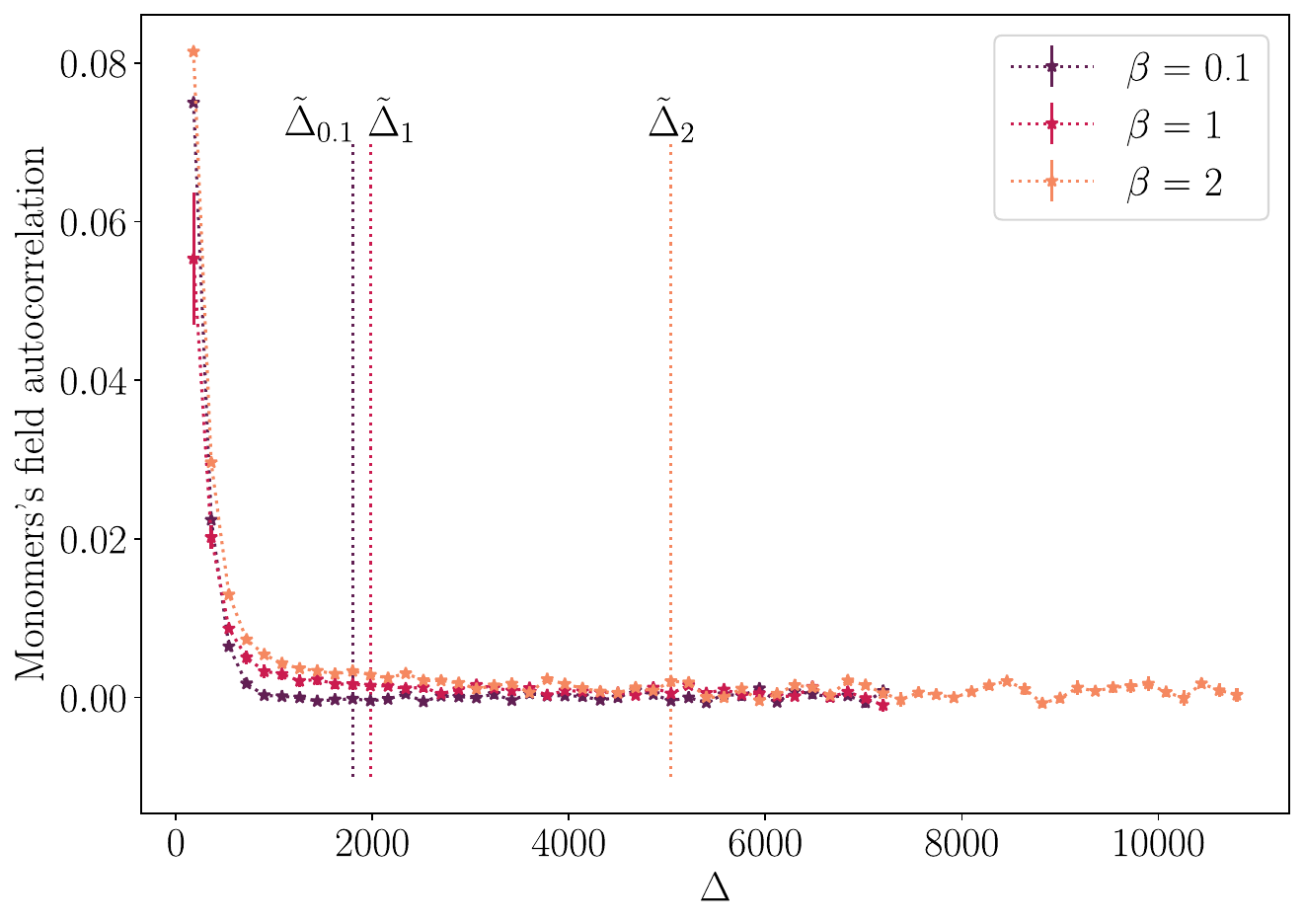}}\\
    \caption{Autocorrelation analysis for simulations run at different inverse temperatures $\beta$. 
    (a) Backbone auto-correlation and (b) monomer field autocorrelation as a function of Markov chain elements  $\Delta$. 
    The vertical dotted lines show the choice of $\tilde{\Delta}_\beta$ for $\beta \in \{0.1,1,2\}$, reported in Tab.~\ref{tab:numerical_details} and used to estimate the distributions of $|\mathbf{i}_\mathrm{i}-\mathbf{i}_\mathrm{f}|$, of the chain's length $\ell$, and of the abundance of $H$ and $P$ residues.
    }
    \label{fig:autocorrelation_time}
\end{figure}

To increase the statistics, we average the autocorrelation function over  $N_\mathrm{mc}$ independent MC trajectories.
The results are reported in Fig.~\ref{fig:autocorrelation_time}.
For practical purposes, we consider configurational states to have reached maximum decorrelation after $\tilde{\Delta}_\beta$ MC steps (see the vertical dotted lines in Fig.~\ref{fig:autocorrelation_time} and numerical values in Tab. \ref{tab:numerical_details}). 
All results reported in the main text are derived by averaging over decorrelated configurations,  sampled from  $N_\mathrm{mc}$ independent Markov chains. \\

\begin{figure}[t!]
    \centering
    \includegraphics[width=\linewidth]{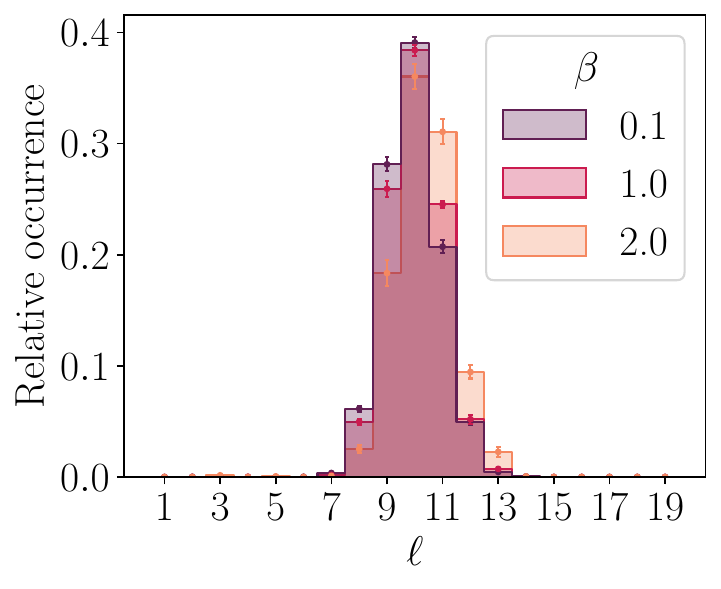}
    \caption{Chain-length distribution for various values of the inverse temperature ($\beta \in \{0.1, 1, 2\}$). Samples are drawn from ensembles where the chain's end-points are delocalized on the lattice. 
    At low temperatures, where entropic effects are small and chemical non-covalent interaction plays a dominant role, the system prefers longer chains.   
    As temperature increases, the chain-length distribution becomes peaked more closely to the value favoured by the chemical potential of $l_0=9$.  
    }
    \label{fig:chain_length_analysis}
\end{figure}

\textbf{Chain length analysis:}
The length distributions for several values of the inverse temperature $\beta$ and parameters detailed in Tab.~\ref{tab:numerical_details} are reported in Fig.~\ref{fig:chain_length_analysis}. For these parameter values, the expected average chain length when neglecting chemical interactions is $l_0=9$. 
At low temperatures, where entropic contributions to the statistical properties of the ensemble are small, we expect the sampling of low-energy configurations to be enhanced. As non-covalent interactions among monomers can lower the protein's energy, and since the possible number of such interactions increases with the chain length, we expect to sample longer polymeric chains at small temperatures. 
Indeed, this is what we observe in our simulations. 
As the temperature increases, the average chain length tends towards the value of $l_0=9$ favored by the chemical potential. 
In all considered cases, thanks to the large value of $M=120$ considered, the chain-length distribution is rather strongly peaked around the average. 

 \section{\label{subsec:SI:determinant_open_chain} Determinant of chain configurations}
In this section, we prove Eq.~\eqref{eq:appendix:determinant_open_chain} of the main text.
The general form of $T(\Gamma_\mathrm{SA})$, where $\Gamma_\mathrm{SA}$ is the binary field associated with a $\ell$--long self-avoiding chain and the parameters are chosen as in Eq.\eqref{choice}, is
\be
    T(\Gamma) 
    = \begin{pNiceArray}{c | c}[margin]
        B_{(\ell+1)} & 0 \\\hline
        0 & D_{(N-\ell-1)}
    \end{pNiceArray}\,,
    \label{eq:open_configuration_matrix}
\ee
where, $D_{(N-\ell-1)}$ is a square diagonal matrix with all entries set to $4\bar{g}^2$ and $B_{(\ell+1)}$ takes the form
\be
    B_{(\ell+1)} = 
    \bar{g}^2
    \begin{pNiceArray}{c c c c c}[margin]
    \color{magenta} 3\color{black} & -1 & \cdot & \cdot & \cdot \\ 
          -1 & 2 & -1 & \cdot &\cdot \\
          \cdot      & -1 & \Ddots & \Ddots & \cdot \\
          \cdot & \cdot   & \Ddots & 2 & -1 \\
          \cdot  & \cdot & \cdot & -1 & \color{cyan} 3\color{black}            
    \end{pNiceArray}_{(\ell+1)\times (\ell+1)}\hspace{-0.7cm}\,.
\label{eq:band_matrix}
\ee 
With this,
\be 
    \det T(\Gamma) = (4\bar{g}^2)^{N -(\ell+1)} \, \det B_{(\ell+1)}\,.
\ee 
To compute the determinant of $B_{(\ell+1)}$, we recall a general result concerning the determinant of block matrices \cite{Abadir2005Matrix}:
\NiceMatrixOptions{cell-space-limits = 1pt}
\be 
\hspace{-0.4cm}
\det
\begin{pNiceArray}{c | c}[margin]
    \Block{1-1}{A} & \Block{1-1}{C}  \\\hline
    \Block{1-1}{D} & \Block{1-1}{B}  
    \label{eq:determinant_block}
\end{pNiceArray} 
= \det A \cross \det\left( A - {\color{black}C} \, B^{-1}\,{\color{black} D}\right)\,.
\ee 
To exploit this result, it is convenient to adopt a change of basis such that the transformed $B^\prime_{(\ell+1)}$ acquires the block-diagonal form
\be
\hspace{-0.5cm}
B^\prime_{(\ell + 1)} = \bar{g}^2
\begin{pNiceArray}{cc|cccc}[margin]
3 & 0 & 0 &\Cdots &   0 & -1 \\
0 & 3 & -1& 0   &\Cdots  & 0  \\\hline 
0 & -1& \Block{*-*}<\Large>{X_{(\ell-1)}} \\
\Vdots & 0\\
0 & \Vdots \\
-1 & 0 
\end{pNiceArray}_{(\ell+1)\times (\ell+1)} \hspace{-1cm}\,,
\ee 
where the matrix $X_{(\ell-1)}$ is a Toepliz matrix having the structure
\NiceMatrixOptions{cell-space-limits = 1pt}
\be
\hspace{-0.5cm}
    X_{(\ell - 1)} = 
    \begin{pNiceArray}{c c c c c c}[margin]
    2  & -1     & 0         & 0         & 0     &0 \\
    -1 & 2      &  -1       & 0         & 0     & 0\\
    0  & -1     & \Ddots    & \Ddots    & 0     & 0\\
    0  & 0      & \Ddots    & 2         & -1    & 0\\
    0  & 0      & 0         & -1        &  2    & -1\\
    0  & 0      & 0         & 0         &  -1   & 2 \\
    \end{pNiceArray}_{(\ell-1)\times(\ell -1)}\hspace{-1cm}\,.
    \label{eq:toeplitz}
\ee
The determinant of $X_{(\ell-1)}$ can be computed through successive applications of Eq.~\eqref{eq:determinant_block} and is found to have the simple expression
\be 
\det X_{(\ell-1)} = \ell\,,
\ee
so that, by applying once more Eq.~\eqref{eq:determinant_block}, the determinant of $B^\prime_{(\ell+1)}$ becomes 
\begin{equation} \begin{split}
    &\det B^\prime_{(\ell+1)} = (\bar{g}^2)^{\ell+1}\det X_{(\ell-1)}\times \\
    &\; \times \det \mqty(3 - (X_{(\ell-1)}^{-1})_{{\color{black}\ell-1,\ell-1}} & - (X_{(\ell-1)}^{-1})_{{\color{black}\ell-1, 1}}\\
    - (X_{(\ell-1)}^{-1})_{{\color{black}1,\ell-1}} & 3-(X_{(\ell-1)}^{-1})_{{\color{black}1, 1}})\,.
    \label{eq:determinant_intermediate}
\end{split}\end{equation}
The column vectors of the inverse matrix needed to evaluate Eq.~\eqref{eq:determinant_intermediate} have the simple form
\begin{align}   
    &\left(X^{-1}_{(\ell-1)}\right)_{k,1} = \frac{\ell - k}{\ell} \label{eq:first_eigenvector}\\
    &\left(X^{-1}_{(\ell-1)}\right)_{k,\ell-1} = \frac{k}{\ell} \qc k = 1,\dots, \ell-1\,. 
    \label{eq:second_eigenvector}
\end{align}   
To check Eqs.~(\ref{eq:first_eigenvector},\ref{eq:second_eigenvector}), it is sufficient to express the $j$-th row of $X_{(\ell-1)}$ as
\begin{equation} 
    \left(X_{(\ell-1)}\right)_{j \cdot} = \begin{cases}
        2\delta_{1,k} - \delta_{2,k} &\qif j=1\,,\\
        2\delta_{j,k} - \delta_{j,k+1} - \delta_{j+1,k} &\qif 1<j<\ell-1\,, \\
        2\delta_{\ell-1,k} - \delta_{\ell-2,k} &\qif j=\ell-1\,, 
    \end{cases}
\end{equation}
so that, plugging Eq.~\eqref{eq:first_eigenvector} in the expression $\pqty{X_{(\ell-1)}}_{j,k} \pqty{X_{(\ell-1)}^{-1}}_{k,1}$, we obtain
\begin{equation}
\begin{split}
    &\sum_k \left(X_{(\ell-1)}\right)_{jk} \frac{\ell-k}{\ell} = \\ 
    &\quad = 
    \begin{cases}
        \frac{2(\ell-1)}{\ell} - \frac{\ell-2}{\ell} = 1 &\qif j = 1\\
        \frac{2(\ell-j)}{\ell} -\frac{\ell-j-1}{\ell} -\frac{\ell-j+1}{\ell} = 0 &\qif 1<j<\ell-1\\
        \frac{2(\ell - (\ell-1))}{\ell} - \frac{\ell-(\ell-2)}{\ell} = 0 & \qif j=\ell-1\,,
    \end{cases} \\
    &\quad = \delta_{j,1}\, 
\end{split}\end{equation}
confirming that Eq.~\eqref{eq:first_eigenvector} is correct. 
To verify that Eq.~\eqref{eq:second_eigenvector} correctly reproduces $\pqty{X_{(\ell-1)}}_{j,k}\pqty{X_{(\ell-1)}^{-1}}_{k,\ell-1} = \delta_{j,\ell-1}$, we evaluate
\begin{equation}\begin{split}
    &\sum_k \pqty{X_{(\ell-1)}}_{j,k} \frac{k}{\ell} = \\
    &\quad = 
    \begin{cases}
        \frac{2}{\ell} - \frac{2}{\ell} = 0 &\qif j=1 \\
        \frac{2 j}{\ell} - \frac{j+1}{\ell} - \frac{j-1}{\ell} = 0 &\qif 1<j<\ell-1\\
        \frac{2(\ell-1)}{\ell} - \frac{\ell-2}{\ell} = 1 &\qif j=\ell-1
    \end{cases} \\
    &\quad = \delta_{j,\ell-1}\,.
\end{split}\end{equation}
Since $B_{(\ell+1)}$ and $B^\prime_{(\ell+1)}$ are related by a change of basis, they share the same determinant, and we finally obtain
\begin{equation}\begin{split}
\det B_{(\ell+1)} &= (\bar{g}^2)^{\ell+1}\;\ell \;
\det\mqty(2 - \frac{1}{\ell} & -\frac{1}{\ell} \\ 
-\frac{1}{\ell} & 2-\frac{1}{\ell}) \\
&= 4(\bar{g}^2)^{\ell+1}(\ell+1)\,.
\label{eq:determinant_open_chain}
\end{split}\end{equation}
Thus,
\be 
    \det T(\Gamma) = 4^{N-\ell} \, g^{2N} \, (\ell+1).
\ee 
\begin{figure}[t!]
\centering
	\includegraphics{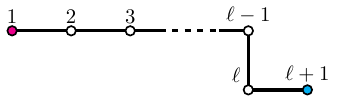}
    \caption{Open chain configuration consisting of $\ell$ edges and $\ell+1$ nodes. 
            In this picture, we omit the representation of the remaining $(N-\ell-1)$ nodes that do not belong to the chain, whose contribution to the $T(\Gamma)$ matrix in Eq.~\eqref{eq:open_configuration_matrix} is embodied by the diagonal square sub-matrix $D_{(N-\ell-1)}$.
            The contribution of the first $\ell+1$ sites belonging to the chain is reflected by the square sub-matrix $B_{(\ell+1)}$ specified in Eq.~\eqref{eq:band_matrix}, 
            where the entries marked in magenta and blue are associated to the extremes of the chain (marked with the same colors in the present figure).}
    \label{fig:open_chain_polymer_configuration}

\end{figure}

\section{Additional figures}

In this section, we report figures that are useful for understanding how Eq.~\eqref{flux} ensures the constraints dictated by the chain's topology and how chain deformation operators $\hat C_\square$ (Eq.~\eqref{eq:wilson_loop}) and end-point displacement operators $\hat D_{\ii\jj}$ (Eq.~\eqref{eq:Ddef}) act on lattice states.
\begin{figure}[h]
\centering
\includegraphics{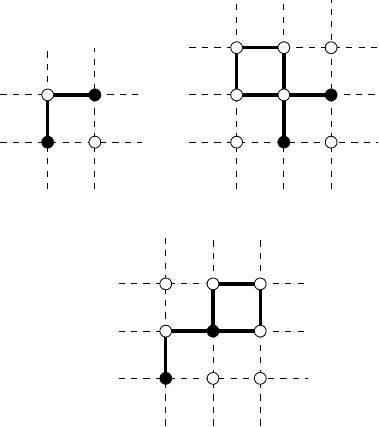}
    \caption{Example of continuous polymer configurations satisfying Eq.~\eqref{flux} of the main text at every vertex. Indeed, the configuration of each vertex is locally among the allowed ones illustrated in Fig.~\ref{fig:physical_sector}.}    
    \label{fig:physical_sector_global}
\end{figure}
\begin{figure}[H]
\centering
\includegraphics{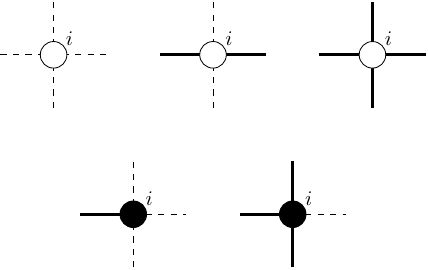}
    \caption{Configurations satisfying the eigenvalue Eq.~\eqref{flux} of the main text at vertex $i$. 
    Black circles represent end-points with a topological charge present, described by $\Gamma_i=1$, while 
    white circles represent regular points along the chain backbone, described by $\Gamma_i = 0$. 
    Solid lines represent active bonds ($\Gamma_\ell = 1$) and dashed lines denote inactive bonds ($\Gamma_\ell = 0$). 
    These configurations can be seen as those permitted by a local symmetry (Gauss' law) encoding flux conservation.}
    \label{fig:physical_sector}
\end{figure}
\begin{figure}[H]
\centering
\includegraphics{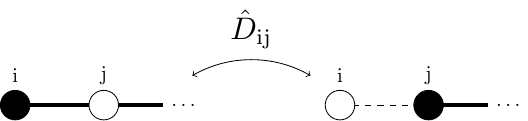}
    \caption{Action of the diffusion operator for the chain's endpoints, as given in Eq.~\eqref{eq:Ddef}. A topological charge is moved along one link, whose active/inactive state is adjusted such that the vertices assume only configurations (illustrated in Fig.~\ref{fig:physical_sector}) that are allowed by the Gauss' law.}
    \label{fig:diffusion_operator}
\end{figure}

\section{\label{sec:SI:sampling_maximally_compact_configurations}Sampling Maximally Compact Configurations with LGT}

In this section, we showcase an example that illustrates why the LGT approach is particularly suited to sample  maximally compact polymer configurations. 

\begin{figure}[t!]
\centering
\includegraphics{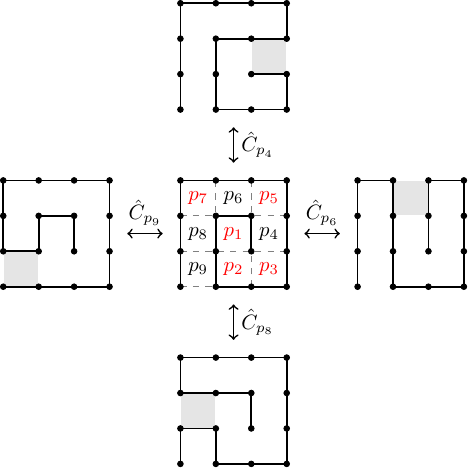}
    \caption{By acting with a single deformation operator (corresponding to plaquettes highlighted in gray), it is possible to derive 4 other Hamiltonian paths from the spiral configuration depicted in the center.}
    \label{fig:Hamiltonian_paths_3_3}
\end{figure}

 We consider the problem of generating  new compact self-avoiding structures by acting with chain deformation operators $\hat C_p$ on the specific Hamiltonian path  state shown at the center of Fig.~\ref{fig:Hamiltonian_paths_3_3}.
This action yields a new viable compact structure when it acts on 4 out of the 9 plaquettes. The rejected moves are those for which $\hat C_p$ acts on a plaquette that contains a path turn, i.e., those highlighted in red in Fig.~\ref{fig:Hamiltonian_paths_3_3}. Hence,  when such a spiral-shaped Hamiltonian path is embedded in a larger two-dimensional lattice of size $N$, the chain deformation moves have a larger success rate,  $\frac{(N-2)^2}{(N-1)^2}$, thus even approaching 100\% efficiency in the thermodynamic limit.

The efficiency of the LGT moves is not surprising, since the action of the chain-deformation operator generates chain rearrangements that resemble those performed in graph-theory-based algorithms for sampling Hamiltonian paths \cite{hamiltonianpath1,hamiltonianpath2} and state-of-the-art MC schemes for sampling maximally compact structures \cite{kineticpath1,kineticpath2}.  
However,  the LGT is defined to tolerate small fluctuations in the chain length. Hence, its generated compact structures are not always Hamiltonian paths. In particular, the simulations concerning homopolymers in a poor solvent reported in the main text have site occupancy of above about $98\%$.

\section{\label{sec:SI:status_quantum_simulation_LGTs}Quantum simulations of LGTs}

In this work, we have reformulated the statistical mechanics of heteropolymers in terms of a LGT. In principle, this encoding paves the way towards resorting to quantum-computing hardware to sample heteropolymer configurations efficiently. 
It is therefore useful to briefly review the recent progress in the simulation of LGTs on quantum-computing facilities.

{\bf Survey of applications of quantum computing to LGTs. }
The earliest uses of quantum computing to simulate gauge theories concerned simple two-site-lattice models.  In particular, Refs.~\cite{Martinez2016} and \cite{Klco2018}, implemented a truncated lattice Schwinger model with $U(1)$ gauge symmetry on a trapped-ions platform and, respectively, on an IBM superconducting quantum chip. 
In both cases, the Gauss law was used to reduce the number of degrees of freedom that needed explicit representation on the quantum simulator. 
An alternative approach to eliminating redundant degrees of freedom is to design intrinsically symmetric quantum platforms. 
For instance, Ref.~\cite{Schweizer2019} demonstrated that a mixture of ultra-cold atoms can be tuned to exhibit genuine $\mathbb{Z}_2$ symmetry.
The same lattice gauge theory was simulated by resorting to a two-species atomic mixture, where matter and gauge degrees of freedom were mapped onto specific atomic entities \cite{Mil2020}. 
A classical LGT with $U(1)$ symmetry in a few-site system was implemented in purpose-built electric circuits \cite{Riechert2022}. 

In recent years, various experiments on different platforms have achieved quantum simulations of LGTs on scales that permit the study of interesting many-body phenomena. 
In Ref.~\cite{Kokail2019}, the ground state of a truncated lattice Schwinger model was simulated using a hybrid quantum-classical algorithm running on an analog quantum co-processor based on trapped ions. 
In Ref.~\cite{Bernien2017}, it was shown that a 51-atom Rydberg platform exhibits $\mathbb{Z}_2$, $\mathbb{Z}_3$, or $\mathbb{Z}_4$ order depending on the laser drivings applied to the atoms, an experiment that was later interpreted in terms of a gauge theory \cite{Surace2020}.
The rigorous mapping between Rydberg-blockaded states and states belonging to the quantum link model of the $U(1)$ LGT, was exploited in Ref.~\cite{Gonzalezcuadra2024} to simulate relevant dynamical processes such as string breaking mechanisms. 
Similar progress has been made in optical-lattice-based setups. In Ref.~\cite{Yang2020} a 71-site-Bose-Hubbard quantum simulator was developed, where Coleman's phase transitions was studied, and Refs.~\cite{Zhou2022, Wang2023} experimentally investigated the thermalization and confinement dynamics of a $U(1)$ lattice gauge model. 
Significant progress has been made in simulating hadron dynamics and in preparing the vacuum state of the truncated lattice Schwinger model using superconducting platforms \cite{Farrell2024PRX, Farrell2024PHYSREVD}. The same technology was leveraged to simulate the dynamics of charges in the $\mathbb{Z}_2$ LGT \cite{Cochran2024} as well as confinement mechanisms \cite{Mildenberger2025}.

Several groups are also involved in experimentally realizing systems that exhibit non-Abelian symmetries, extending to higher dimensions, and including dynamical fermions~\cite{Zohar2021}. For example,  Ref.~\cite{Ciavarella2022} addressed the vacuum of a Yang-Mills $SU(3)$ LGT with the use of a variational quantum encoder protocol performed on superconducting chips. Experiments involving QCD systems in $(1+1)$ dimensions have also been implemented using trapped-ions devices \cite{Farrell2023}.
 For a recent review that includes further promising research directions see, e.g.,~\cite{Wiese2014,Zohar2015,Dalmonte2016,Banuls2019,Banuls2020,Klco2022,Zohar2021,Aidelsburger2021,DiMeglio2024,Halimeh2023}.

{\bf Perspective on future application of quantum computing to our polymer LGT}. 
The present LGT formulation of polymer physics resorts to the local $\mathbb{Z}_2$  invariance,  a simple symmetry to implement on a quantum computer or simulator.  
The key technical development that is needed to sample single-chain configurations is the quantum encoding of the determinant $\det(T(\Gamma))$ entering Eq.~\eqref{eq:partition_function} of the main text. To this end, a strategy worth exploring would be to first design a gate-based circuit that takes care of evaluating the determinant associated with an input binary field configurations and then implementing the elevation to the $M$-th power, by relying on block-encoding and quantum singular value transformation techniques~\cite{QSVT}. Such an approach would scale only polynomially in the number of qubits and, importantly,  would not depend on the number of fermions $M$. Even if one would simulate the fermions explicitly, the computational cost would still be linear in volume and in $M$. Another strategy well suited for Noisy Intermediate Scale Quantum (NISQ) devices would be to resort to quantum approximate optimization Ansatz (QAOA) techniques \cite{Farhi2014,Blekos2024}.   In such an approach, the cost function to be evaluated in classical postprocessing would be the one given in   Eq.~\eqref{eq:partition_function},  whereas the variational quantum circuit would only contain the native gates acting on the qubits representing the polymer. By optimizing the circuit parameters with a classical feedback loop, the circuit would sample only loop-free low-energy configurations. 

\nocite{Abadir2005Matrix}
\nocite{Martinez2016}
\nocite{Klco2018}
\nocite{Schweizer2019}
\nocite{Mil2020}
\nocite{Riechert2022}
\nocite{Kokail2019}
\nocite{Bernien2017}
\nocite{Surace2020}
\nocite{Gonzalezcuadra2024}
\nocite{Yang2020}
\nocite{Zhou2022}
\nocite{Wang2023}
\nocite{Farrell2024PRX}
\nocite{Farrell2024PHYSREVD}
\nocite{Cochran2024}
\nocite{Mildenberger2025}
\nocite{Zohar2021}
\nocite{Ciavarella2022}
\nocite{Farrell2023}
\nocite{Wiese2014}
\nocite{Zohar2015}
\nocite{Dalmonte2016}
\nocite{Banuls2019}
\nocite{Banuls2020}
\nocite{Klco2022}
\nocite{Aidelsburger2021}
\nocite{DiMeglio2024} 
\nocite{Halimeh2023}
\nocite{QSVT}
\nocite{Farhi2014}
\nocite{Blekos2024}

\end{document}